\documentclass[aps,pre,a4paper,10pt,twoside,twocolumn,floatfix]{revtex4-2}

\usepackage{amsmath,amssymb,bm}
\usepackage{mathptmx,microtype}

\usepackage[colorlinks,linkcolor=blue,citecolor=blue,urlcolor=blue]{hyperref}
\usepackage{graphicx}

\newcommand{\mku}{\ensuremath{\mkern1mu}}

\newcommand{\ie}{i.{\kern1pt}e.}
\newcommand{\iid}{i.{\kern1pt}i.{\kern1pt}d.}
\newcommand{\eps}{\ensuremath{\varepsilon}}
\newcommand{\abs}[1]{\left|#1\right|}
\renewcommand{\mid}{\ensuremath{\mkern2mu|\mkern2mu}}

\DeclareMathOperator{\PP}{\mathbb{P}}
\DeclareMathOperator{\EE}{\mathbb{E}}

\begin{document}

\title{Records, drift, and the longest increasing subsequence of biased Gaussian random walks}

\author{J.~Ricardo G. Mendon\c{c}a\hspace{1pt}\href{https://orcid.org/0000-0002-5516-0568}{\includegraphics[trim=-5 0 0 0, scale=0.20]{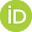}}}
\email{jricardo@usp.br}
\affiliation{Escola de Artes, Ci\^{e}ncias e Humanidades, Universidade de S\~{a}o Paulo, 03828-000 S\~{a}o Paulo, Brazil}

\begin{abstract}
The longest increasing subsequence (LIS) of a random walk has been studied mainly for zero-mean, symmetric step increments. We numerically investigate the LIS of biased Gaussian random walks, with unit-variance increments and positive drift $\mu_{p} = \Phi^{-1}(p)$, where $p = \PP(\xi>0)$. In contrast with the symmetric case, we find that for every fixed $p>1/2$ the mean LIS length grows linearly, $\langle L_{n}(p)\rangle \sim a(p)n$, with $a(p)$ increasing from $0$ at $p=1/2$ to $1$ as $p \to 1$. The record count is also linear, with coefficient $\lambda(p)$ fixed by Spitzer's formula for the ascending ladder epoch, and the LIS becomes increasingly aligned with this record skeleton as $p$ grows. At the symmetric point $p=1/2$, the record skeleton collapses to the Sparre Andersen $\sqrt{n}$ scale, while the LIS returns to the finite-variance $\sqrt{n}\mku\log{n}$ regime. Near this limit the record rate has the closed-form small-drift slope $\lambda(\mu_{p}) \simeq \sqrt{2}\,\mu_{p}$, whereas the excess $a(\mu_{p})-\lambda(\mu_{p})$ vanishes more slowly than linearly in the drift, although our data do not resolve a single power law. The empirical distribution of $L_{n}$ also changes across this point, from lognormal-like at $p=1/2$ to Gaussian-like for every sampled $p>1/2$.
\end{abstract}


\maketitle


\section{Introduction}\label{intro}

The classic longest increasing subsequence (LIS) problem, first proposed by Ulam in the early 1960s, asks for the properties of the increasing subsequences $\pi_{i_{1}} < \cdots < \pi_{i_{k}}$, $1 \leq i_{1} < \cdots < i_{k} \leq n$, of maximal length $k$ in random $n$-permutations $\pi$ \cite{ulam,hammersley}. The LIS problem for random permutations proved remarkably rich, and its study led to developments across subjects ranging from group representation theory to random matrix theory, algebraic combinatorics, determinantal point processes, and exactly solvable models of statistical mechanics. We refer the reader to \cite{bdj99,aldous,patience,romik,gorin} for comprehensive expositions and for further references.

Recently, the LIS problem for random walks has also begun to receive attention \cite{angel,pemantle}. Several recent numerical studies have investigated the properties of the LIS
of random walks \cite{lisjpa,hartmann,lispre,fatlis,weakheavy}. These studies confirmed the few existing rigorous results and identified candidate scaling forms for the leading asymptotic behavior of the LIS length of random walks with short- and heavy-tailed step increments, as well as for its full distribution function and large-deviation rate function. All results obtained so far, however, refer to the LIS of symmetric random walks. The characterization of the LIS of biased random walks remained unexplored. Biased random walks model time series with a directional trend; the LIS then probes monotone structure in a correlated, nonstationary sequence, a setting relevant to the analysis of real-time and large-scale data streams \cite{liben,gopalan2010}. While analytical results are not available, the relevant observables are well-defined, and a numerical investigation offers a first step.

We study the LIS of biased random walks, in which the step increments are Gaussian with unit variance and a positive mean $\mu_{p}$ set by the bias parameter $p = \PP(\xi>0) \geq 1/2$. The choice of Gaussian increments isolates the effect of asymmetry without introducing heavy-tail effects, in a setting where the symmetric case is well-understood. As we shall see, the leading asymptotic behavior of the LIS length differs qualitatively from the symmetric one already at vanishing asymmetry, and an asymptotic exponent in the sense of the symmetric scaling laws is no longer the informative object. In this case one should appeal to the theory of records of biased random walks, in which the universal Sparre Andersen behavior at $p=1/2$ is replaced by a linearly growing record skeleton with which the LIS aligns~\cite{majumdarziff,gms,wbk,msw}.

We find that any positive bias makes the mean LIS length grow linearly in $n$ throughout $1/2 < p < 1$, so the object of interest is the growth rate $a(p)$ rather than a scaling exponent. This contrasts with the heavy-tailed case, where the LIS grows sublinearly with an exponent that varies continuously with the tail index \cite{weakheavy}. The mean record count is also linear, with rate $\lambda(p) \leq a(p)$, and the gap $a(p)-\lambda(p)$ measures the part of the LIS not carried by the record skeleton. We also identify the exact small-drift slope of the record rate, $\lambda(\mu_{p})\simeq\sqrt{2}\,\mu_{p}$ (Appendix~\ref{app:smalldrift}), used previously in this setting only as a numerical estimate~\cite{wbk}; it is the record-rate form of a classical ladder-height result~\cite{changperes}. The symmetric point $p=1/2$ is a singular limit of this linear regime: as the drift vanishes, $a(p)-\lambda(p)$ goes to zero more slowly than linearly, with no single power law resolved over the range we sample. The distribution of $L_{n}$ changes abruptly across the same point: lognormal-like at $p=1/2$, as reported for symmetric walks \cite{weakheavy}, and consistent with Gaussian fluctuations for every sampled $p>1/2$.

This paper is organized as follows. In Section~\ref{sec:known} we review the main facts about the LIS of symmetric random walks and the role of records, fixing notation. Section~\ref{sec:normal} defines the biased Gaussian walk and the observables we measure. Section~\ref{sec:linear} characterizes the linear regime, reports the coefficient $a(p)$, identifies the record-density coefficient $\lambda(p)$ with the known closed form~\cite{msw}, and documents the empirical transition in the distribution of $L_{n}$ across $p=1/2$. In Section~\ref{sec:records} we interpret the gap $a(p)-\lambda(p)$ as the LIS contribution from non-record interpolations between consecutive records of the walk. Section~\ref{sec:symlimit} treats $p=1/2$ as a singular limit of the linear regime and identifies the drift crossover scale $\mu_{c}(n)$ at which the symmetric prediction breaks down. Finally, in Section~\ref{sec:summary} we summarize and discuss our results and indicate directions for further research involving distributions of infinite variance. Appendix~\ref{app:smalldrift} derives the small-drift limit of the record rate.


\section{Background}
\label{sec:known}

\subsection{The LIS of random walks}
\label{sec:known:lis}

An $n$-step random walk is a sequence $(X_{1}, \dots, X_{n})$ of integer or real numbers given by
\begin{equation}
\label{eq:rwalk}
X_{0}=0, \quad X_{k} = X_{k-1}+\xi_{k}, ~1 \leq k \leq n,
\end{equation}
where the step increments $\xi_{k}$ are independent and identically distributed (\iid) random variables according to some probability distribution $\phi(x \mid \bm{\beta})$. The walk positions are correlated according to 
\begin{equation}
\operatorname{Cov}(X_{s},X_{t}) = \min(s,t)\operatorname{Var}(\xi);
\end{equation} 
the LIS problem for random walks differs from the permutation case because the entries of the sequence are not independent. Given a random walk $(X_{1}, \dots, X_{n})$, we are interested in the behavior of its longest strictly increasing subsequence
\begin{equation}
X_{i_{1}} < \cdots < X_{i_{k}}, \quad 1 \leq i_{1} < \cdots < i_{k} \leq n.
\end{equation}
A sequence can have more than one LIS, with different elements but the same maximal length $k$. For continuous distributions of step increments, ties occur with probability zero, so the distinction between weakly ($X_{i_{1}} \leq \cdots \leq X_{i_{k}}$) and strictly increasing subsequences is immaterial. We therefore refer to the LIS without qualification throughout. The LIS of a sequence of $n$ numbers can be calculated in $O(n)$ space and $O(n\log{n})$ time by the patience sorting algorithm \cite{patience,sergei}.

\subsection{Known facts}
\label{sec:knownfacts}

In the mathematical literature, we can find the following two main rigorous results about the LIS of zero-mean ($\EE(\xi)=0$), symmetric random walks. First, when $\phi(x \mid \bm{\beta})$ is of finite variance, it has been proved that for all $\eps>0$ and sufficiently large $n$ the expected LIS length $\EE(L_{n})$ is bounded by
\begin{equation}
\label{eq:finite}
c\sqrt{n} \leq \EE(L_{n}) \leq n^{1/2+\eps}
\end{equation}
for some $c>0$ \cite{angel}; the lower bound follows from the Erd\H{o}s--Szekeres theorem \cite{hammersley,aldous}. For the simple, discrete random walk on $\mathbb{Z}$ with step increments $\pm 1$, where repeated visits to lattice levels create plateaus that admit weakly but not strictly increasing extensions, the lower bound on the expected length of the weak LIS can be improved to $c\sqrt{n}\mku\log{n} \leq \EE(L_{n})$ \cite{angel}. Second, when the distribution of step increments $\phi(x \mid \bm{\beta})$ has infinite variance, the bounds on $\EE(L_{n})$ become
\begin{equation}
\label{eq:infinite}
n^{\beta_{0}-\eps} \leq \EE(L_{n}) \leq n^{\beta_{1}+\eps},
\end{equation}
where $\beta_{0} \simeq 0.690$ is the positive solution of $x+2^{-1-x}=1$ and $\beta_{1} \simeq 0.815$ is obtained from the numerical solution of an implicit equation involving a non-elementary integral \cite{pemantle}; in this case, the exponents are not sharp.

Numerical simulations show that the behavior of $L_{n}$ depends on the increment distribution. When $\phi(x \mid \bm{\beta})$ is of finite variance, it has been found that the LIS of the random walk behaves asymptotically as
\begin{equation}
\label{eq:short}
L_{n} \sim a\sqrt{n}\mku\log{n} + b\sqrt{n}
\end{equation}
with $a \simeq b \simeq 0.36$ \cite{lisjpa,hartmann,lispre}. When, instead, $\phi(x \mid \bm{\beta})$ is heavy tailed, it has been found that
\begin{equation}
\label{eq:heavy}
L_{n} \sim n^{\theta}
\end{equation}
with a varying exponent in the range $0.5 < \theta \lesssim 0.716$ depending on the tail index $\alpha$ according to which $\phi(x \mid \bm{\beta}) \sim \abs{x}^{-1-\alpha}$ as $\abs{x} \to \infty$ \cite{lisjpa,hartmann,lispre,fatlis}. The largest exponent $\theta \simeq 0.716$ was found for the ultra-fat tailed random walk, which has the heaviest possible distribution of step increments \cite{pemantle,fatlis}. More recently, a study of discrete heavy-tailed random walks on the integers \cite{weakheavy} discriminated between (\ref{eq:short}) and (\ref{eq:heavy}) by combining weighted nonlinear least squares with ANOVA-based model selection, confirming the two-regime picture. The same study reports that the bulk of the distribution of $L_{n}$ is well-approximated by a lognormal across all $\alpha$ and $n$ examined, supplanting earlier Gumbel-collapse conjectures \cite{lisjpa,lispre}. The $\log{n}$ correction in (\ref{eq:short}) is rigorously established only for the weak LIS of discrete random walks; whether it reflects an actual property of the LIS for continuous distributions of step increments or is an artifact of the discreteness of simulations remains unclear.

\subsection{The role of records}
\label{sec:known:records}

Define the upper record times of the random walk by 
\begin{equation}
T_{0}=0, \quad T_{k+1}=\inf\{n>T_{k}\colon X_{n} > X_{T_{k}}\},
\end{equation}
and the record count 
\begin{equation}
R_{n} = \max\{k\colon T_{k} \leq n\}.
\end{equation}
Because the values $X_{T_{1}} < X_{T_{2}} < \cdots$ are automatically strictly increasing, any subsequence indexed by record times is a valid increasing subsequence and, clearly,
\begin{equation}
\label{eq:rec}
L_{n} \geq R_{n}
\end{equation}
for every realization. By the Sparre Andersen theorem, the record count of a random walk with a symmetric continuous distribution of step increments is universal in the increment law,
\begin{equation}
\EE(R_{n}) \sim \sqrt{4n/\pi}, 
\end{equation}
and, more strongly, the entire distribution of $R_{n}$ is independent of $\phi(x \mid \bm{\beta})$ for any symmetric continuous $\phi$ \cite{majumdarziff,gms}. In the biased case this universality breaks down. For random walks with a finite-variance jump distribution and a constant positive drift, the record count grows essentially linearly in $n$, 
\begin{equation}
\EE(R_{n}) \sim \lambda n,
\end{equation}
with a rate $\lambda$ that depends on the drift and on the increment distribution. The biased random walk was first analyzed in \cite{wbk}, where the small- and large-drift asymptotics of the record rate were derived. The exact prefactor is known for finite-variance continuous jump distributions with positive drift~\cite{gms,msw}; we give the explicit formula and a renewal-theoretic interpretation via Spitzer's identity in Section~\ref{sec:linear:fwk}. The implication for the LIS is immediate: since $L_{n} \geq R_{n}$, an enhanced record density $\sim\lambda n$ forces $L_{n}/n$ to be bounded away from zero, so $\theta(p) = 1$ for any $p > 1/2$, and the question becomes the value of the leading coefficient $a(p)$. The fraction $R^{\text{LIS}}_{n}/R_{n}$ of walk records that participate in a given LIS provides a diagnostic for distinguishing record-driven from plateau- or fluctuation-driven LIS growth. For biased walks with positive drift this ratio is expected to approach unity, but its rate of approach as a function of $p$ is one focus of the present work.


\section{The biased Gaussian random walk}
\label{sec:normal}

\subsection{Definition and drift decomposition}
\label{sec:normal:def}

The Gaussian $n$-step random walk is the sequence $(X_{1}, \dots, X_{n})$ of real numbers given by (\ref{eq:rwalk}) with the step increments $\xi_{k}$ distributed according to the Gaussian distribution 
\begin{equation}
\phi(x \mid \mu,\sigma^{2}) = \frac{1}{\sqrt{2\pi}\mku\sigma}\exp[-(x-\mu)^{2}/2\sigma^{2}]. 
\end{equation}
Since the LIS depends only on the relative order of the walk values, multiplication of all increments by a positive constant leaves it unchanged; the relevant scale-invariant parameter of the Gaussian family is thus $p = \PP(\xi>0)$, and we set $\sigma^{2}=1$ without loss of generality. The LIS of the symmetric ($\mu=0$) Gaussian random walk was studied in \cite{lisjpa,lispre} and its behavior agrees with (\ref{eq:short}).

\begin{figure}[t]
\centering
\includegraphics[viewport= 0 10 480 460, scale=0.35, clip]{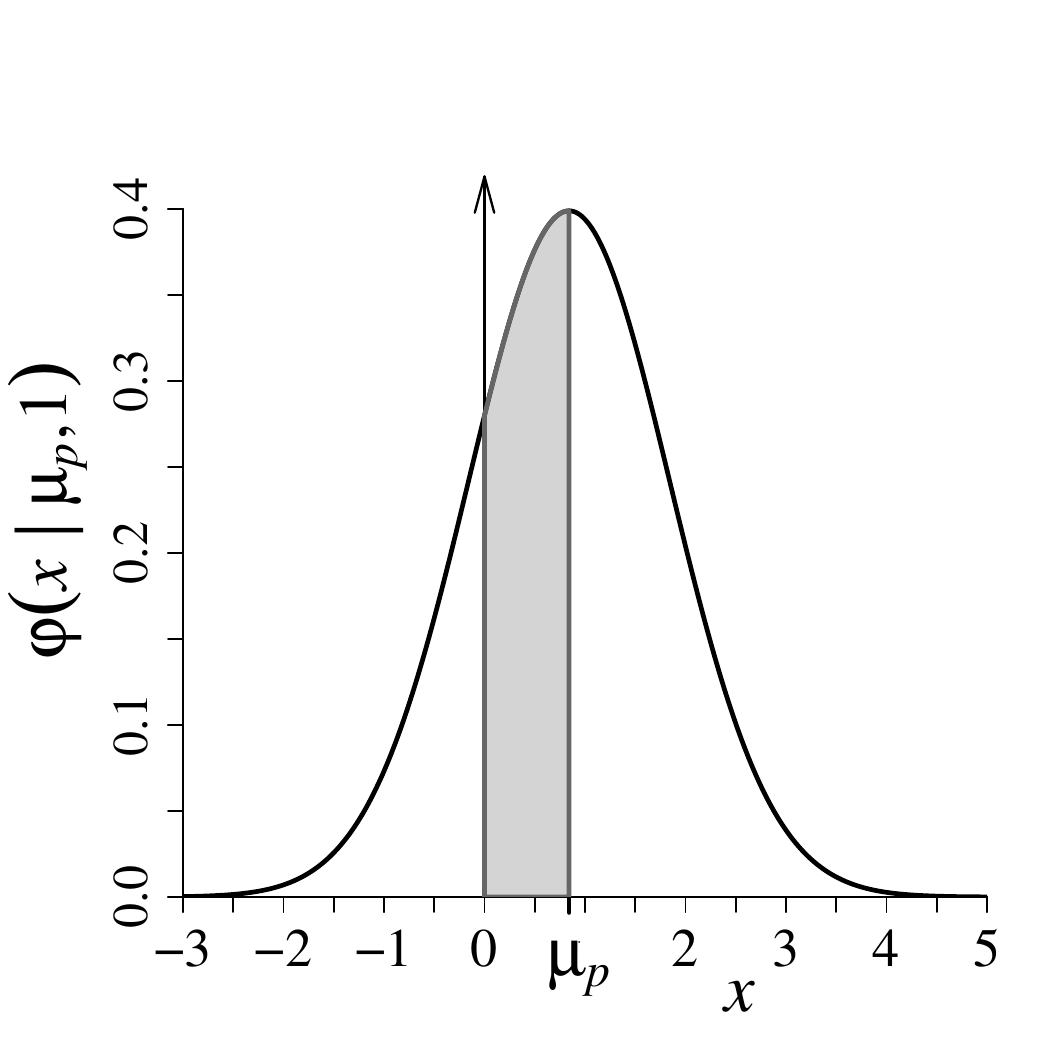}
\caption{Gaussian increment density $\phi(x \mid \mu,\sigma^2)$ with $\sigma^2=1$ and $\mu = \mu_{p} = \Phi^{-1}(p)$, so that $\PP(\xi>0) = p \geq 1/2$. The shaded area has mass $p-1/2$ relative to the centered case $\mu=0$.}
\label{fig:mu}
\end{figure}

We introduce the asymmetry by taking $p \geq 1/2$, obtained by setting $\mu$ to the $p$-quantile of the standard Gaussian distribution (see Figure~\ref{fig:mu}),
\begin{equation}
\label{eq:mu}
\mu = \mu_{p} = \Phi^{-1}(p),
\end{equation}
where $\Phi$ denotes the standard Gaussian cumulative distribution function $\Phi(x) = (1/\sqrt{2\pi})\int_{-\infty}^{x}\exp(-u^{2}/2)\mku du$.

Setting $\mu_{p}>0$ endows the walk with a constant positive drift $\EE(\xi)=\mu_{p}$, and (\ref{eq:rwalk}) decomposes as
\begin{equation}
\label{eq:drift}
X_{k} = \mu_{p}\mku k + W_{k}, \quad W_{k} = \sum_{j=1}^{k}(\xi_{j}-\mu_{p}),
\end{equation}
where $W_{k}$ is a centered Gaussian random walk of unit variance per step. The decomposition (\ref{eq:drift}) is reminiscent of the linear drift model $X_{k}=Y_{k}+ck$ studied in \cite{krug}, with the essential difference that $W_{k}$ in (\ref{eq:drift}) is a correlated walk, whereas $Y_{k}$ there is \iid. The relevant theory here is therefore the records theory of biased random walks \cite{majumdarziff,gms,wbk,msw}, in which any nonzero $\mu_{p}$ produces a record count $R_{n}(p)$ that grows linearly in~$n$, in contrast with the Sparre Andersen universal $\sqrt{4n/\pi}$ behavior at $\mu_{p}=0$. We present the closed-form record rate $\lambda(p)$ and its renewal-theoretic interpretation via Spitzer's identity in Section~\ref{sec:linear:fwk}; combined with the structural inequality $L_{n}(p) \geq R_{n}(p)$, the enhanced record density forces $L_{n}(p)$ to grow at least linearly in $n$.

\subsection{Observables and simulation protocol}
\label{sec:normal:obs}

For each realization of the walk we measure the length $L_{n}(p)$ of the LIS, the record count $R_{n}(p)$, and the number $R_{n}^{\text{LIS}}(p)$ of record times that belong to a recovered LIS. When more than one LIS exists, we use the LIS returned by the patience-sorting algorithm \cite{patience,sergei}. This choice does not affect $L_{n}$, and test runs indicate that it does not affect the sample averages of the overlap observables reported below. These three observables yield the leading-order coefficients
\begin{equation}
\label{eq:coefs}
\begin{split}
a(p) &= \lim_{n \to \infty}\frac{\langle L_{n}(p)\rangle}{n}, \\[1ex]
\lambda(p) &= \lim_{n \to \infty}\frac{\langle R_{n}(p)\rangle}{n}, \\[1ex]
r(p) &= \lim_{n \to \infty}\frac{\langle R_{n}^{\text{LIS}}(p)\rangle}{n},
\end{split}
\end{equation}
that characterize the linear regime (Section~\ref{sec:linear}), together with the diagnostic ratios $R_{n}^{\text{LIS}}/L_{n} \to r(p)/a(p)$ and $R_{n}^{\text{LIS}}/R_{n} \to r(p)/\lambda(p)$ that probe the mechanism by which the LIS is assembled (Section~\ref{sec:records}). Since $R_{n}^{\text{LIS}} \leq \min(L_{n},R_{n})$ and $L_{n} \geq R_{n}$, the bounds $r(p) \leq \lambda(p) \leq a(p)$ follow.

We generate $10\mku000$ independent realizations of the biased Gaussian random walk for each value of the walk length $n$ on the grid $\{1,2,5,10\} \times \{10^{4},10^{5},10^{6}\}$ and for each value of the bias parameter $1/2 \leq p < 1$, with more points near the symmetric point $p=1/2$ and the near-deterministic limit $p \to 1$, where the observables change rapidly. Since $\mu_{p} \to +\infty$ as $p \to 1$, the largest $p$ used is $p=0.999$, for which $\mu_{p} \simeq 3.09$; all observables have saturated to within four decimal places of their $p=1$ limit by $p=0.999$, so this is not an issue. For each realization, the LIS is computed by patience sorting in $O(n\log{n})$ time and $O(n)$ space, the record count by a single running-maximum pass, and the LIS-record overlap by backtracking through the predecessor pointers and counting intersections with the record set in $O(n)$ time and space.


\section{Linear scaling of the LIS}
\label{sec:linear}

\subsection{Effective exponent}
\label{sec:linear:expo}

\begin{figure}[t]
\centering
\includegraphics[viewport= 0 10 480 460, scale=0.40, clip]{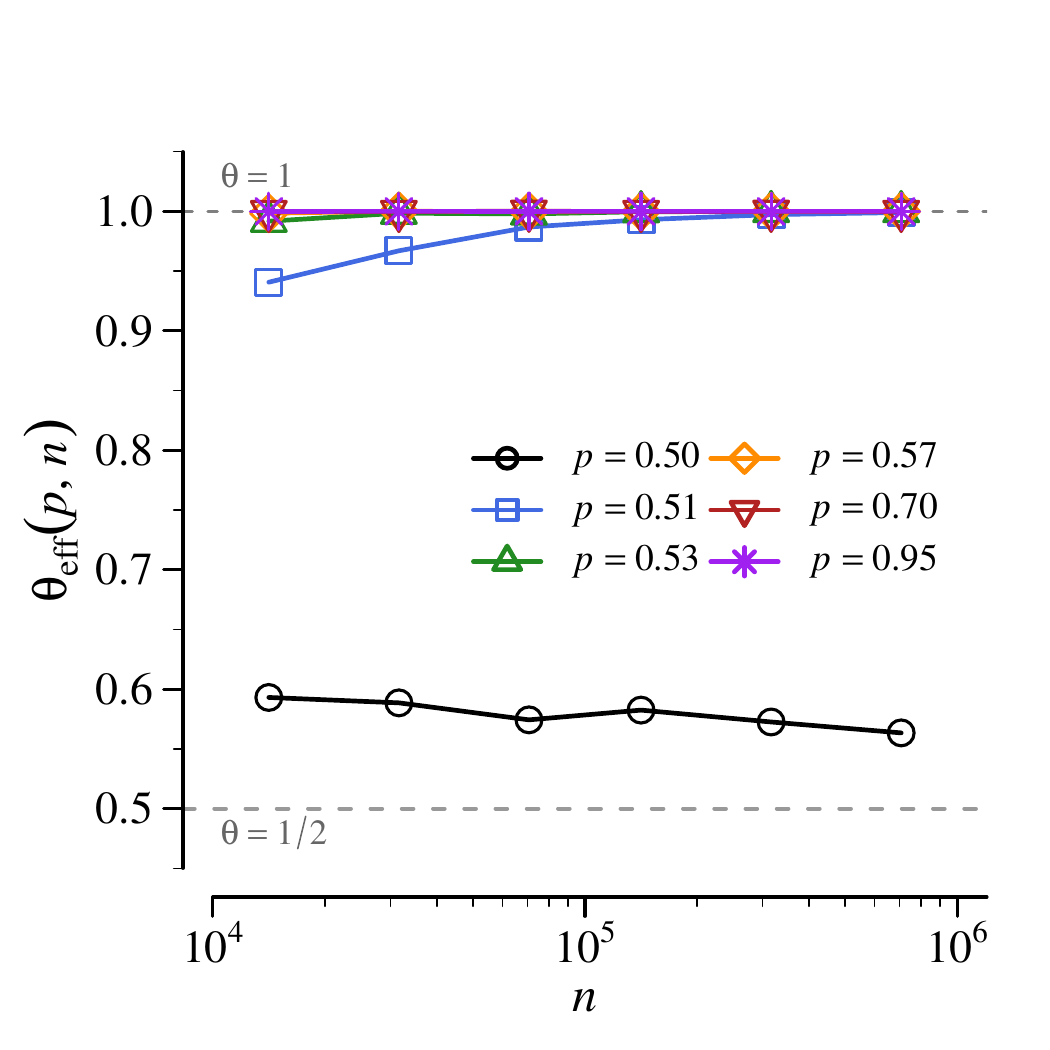}
\caption{Effective exponent $\theta_{\text{eff}}(p,n) = \Delta\log\langle L_{n}\rangle/\Delta\log{n}$, evaluated at the geometric midpoint of consecutive walk lengths. For every sampled $p>1/2$, the curves rapidly approach $\theta=1$. The symmetric case $p=1/2$ remains below unity, consistent with the finite-variance $\sqrt{n}\mku\log{n}$ regime.}
\label{fig:theta_eff}
\end{figure}

The biased Gaussian random walk departs from the symmetric scaling laws (\ref{eq:short})--(\ref{eq:heavy}) already at the smallest asymmetries we consider. Following \cite{weakheavy}, we extract the effective exponent
\begin{equation}
\label{eq:effexp}
\theta_{\text{eff}}(p,n) = \frac{\Delta\log\langle L_{n}(p)\rangle}{\Delta\log{n}}
\end{equation}
from consecutive pairs of walk lengths. The result is plotted in Figure~\ref{fig:theta_eff}. For every sampled $p>1/2$, $\theta_{\text{eff}}(p, n)$ is at or close to $1$. At $p \geq 0.53$ convergence is complete already at $n=10^{4}$, and at $p=0.51$ it is essentially complete by $n=10^{5}$. Only the symmetric case $p=1/2$ remains distinctly below unity, with $\theta_{\text{eff}} \simeq 0.57$ at our smallest $n$, drifting slowly downward toward $1/2$ as $n$ grows, consistent with the symmetric $\sqrt{n}\mku\log{n}$ correction of (\ref{eq:short}). Together with the record lower bound $L_{n}(p) \geq R_{n}(p)$, this confirms that the leading exponent is $\theta(p)=1$ on the entire interval $1/2 < p < 1$. The relevant object is therefore the LIS coefficient $a(p) = \lim_{n \to \infty} \langle L_{n}(p)\rangle/n$, addressed in the next subsection.

\subsection{The linear coefficient $a(p)$}
\label{sec:linear:ap}

\begin{figure}[t]
\centering
\includegraphics[viewport= 0 10 480 460, scale=0.40, clip]{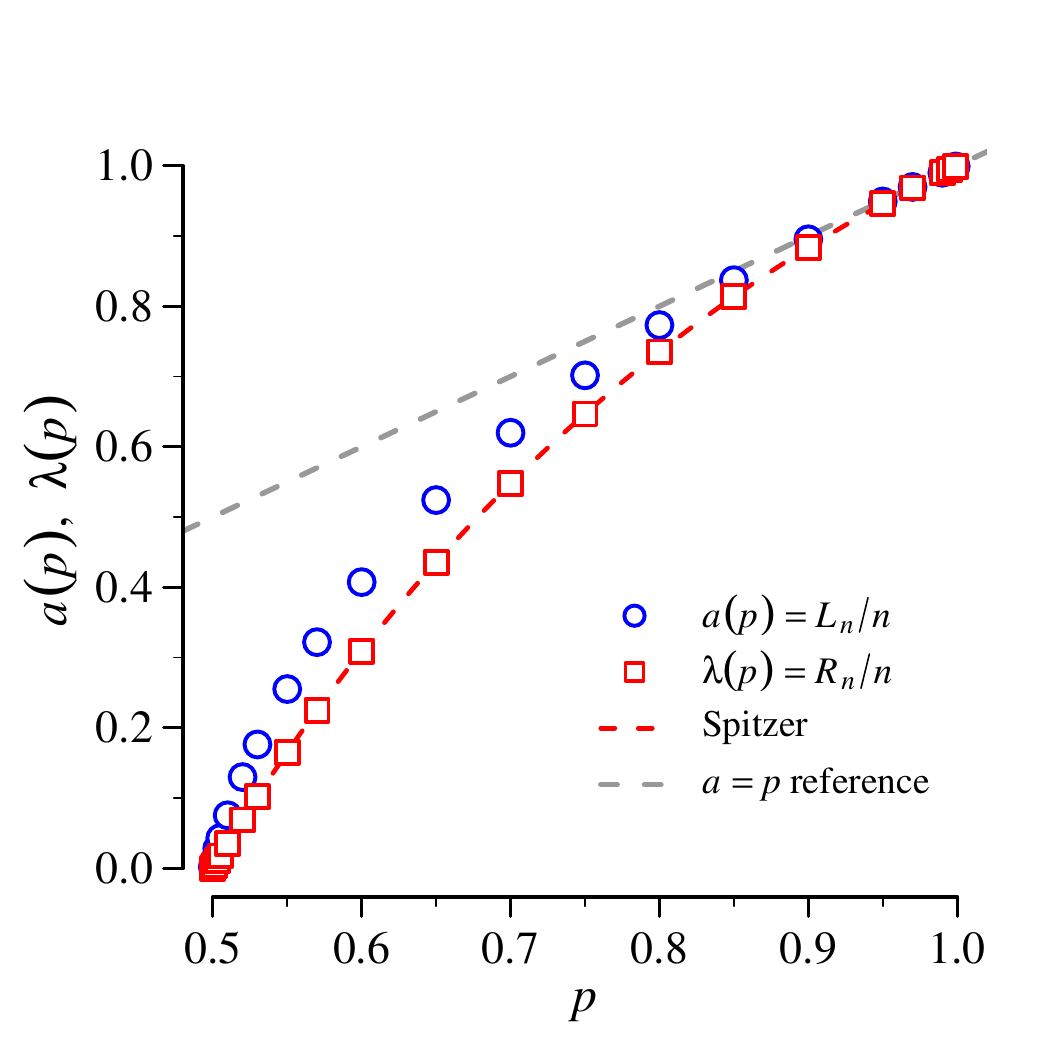}
\caption{Linear coefficients $a(p) = \langle L_{n}\rangle/n$ and $\lambda(p) = \langle R_{n}\rangle/n$ for the biased Gaussian random walk at $n=10^{7}$, with $10^{4}$ samples. The red dashed curve is the theoretical prediction~(\ref{eq:lambdagauss}), and the gray dotted line is the high-drift reference $a=p$. At $p=1/2$, the plotted points are finite-$n$ ratios; the limiting linear coefficients vanish. Standard errors are smaller than the markers.}
\label{fig:a_lambda}
\end{figure}

For any $p>1/2$ the data are consistent with 
\begin{equation}
\langle L_{n}(p) \rangle \sim a(p)\mku n
\end{equation}
to leading order in $n$, with a coefficient $a(p)$ that varies smoothly from $a(1/2)=0$ to $a(1)=1$. The values of $a(p)$ extracted from the largest walk length $n=10^{7}$ are shown in Figure~\ref{fig:a_lambda}, together with the corresponding values of $\lambda(p)$ (see Section~\ref{sec:linear:lambdap}) and the closed-form theoretical prediction for $\lambda$ (see Section~\ref{sec:linear:fwk}).

At large $p$, the data suggest the simple limiting behavior $a(p) \to p$, with the convergence already at the four-decimal level by $p=0.99$. Heuristically, when the drift is large, downward excursions become rare and the positive-increment steps, record times, and LIS elements nearly coincide. Quantitatively, for $\mu_{p} \gg 1$ the probability that a single increment is negative is 
\begin{equation}
\Phi(-\mu_{p}) \sim \frac{e^{-\mu_{p}^{2}/2}}{\sqrt{2\pi}\mku\mu_{p}}. 
\end{equation}
For the record coefficient, the same leading exponential scale appears in the large-drift asymptotics of~\cite{wbk}, recovered from (\ref{eq:lambdagauss}) in Section~\ref{sec:linear:fwk}, giving $\lambda(p)-p \to 0$ exponentially fast as $p \to 1$. The observed behavior of $a(p)-p$ and $r(p)-p$ is consistent with the same exponentially small approach. This explains why the LIS density approaches the fraction $p$ of positive increments, although the positive-increment times $\{k\colon \xi_{k} > 0\}$ do not, in general, form an increasing subsequence of walk values for finite $p<1$. This near-coincidence is also reflected in the saturation of the record-usage ratios at large $p$ (Section~\ref{sec:records:ratios}); the corresponding small-$p$ behavior, where the symmetric $\sqrt{n}\mku\log{n}$ regime is being shed and $a(p)$ lifts off from $0$, remains the unresolved part of the curve and is addressed separately in Section~\ref{sec:symlimit:approach}.

The deficit $D_{n}(p) = n-\langle L_{n}(p)\rangle$ grows linearly as
\begin{equation}
D_{n}(p) = (1-a(p))\mku n + \text{subleading},
\end{equation}
with a subleading correction below our resolution for $p \geq 0.53$, so the deficit carries no information beyond $a(p)$. Finite-size deviations of $\langle L_{n}(p)\rangle/n$ from its $n \to \infty$ value are below $10^{-4}$ already at $n=10^{4}$ and below $10^{-5}$ at $n=10^{6}$; only at $p=0.51$, the smallest asymmetry, is the residual marginally above the noise floor at the smallest sizes before falling below resolution. This is qualitatively consistent with the exponential cutoff of the persistence probability
\begin{equation}
Q(m) = \PP(X_{1}, \dots, X_{m} \leq 0) \sim m^{-3/2}\exp[-(\mu_{p}^{2}/2)m]
\end{equation}
derived in \cite[Eq.~(43)]{msw} for Regime~IV, and our data cannot isolate a separate asymptotic correction to the linear deficit.

\subsection{The record-density coefficient $\lambda(p)$}\label{sec:linear:lambdap}

The record count satisfies 
\begin{equation}
\langle R_{n}(p) \rangle \sim \lambda(p)n
\end{equation}
with a similar smooth coefficient $\lambda(p)$ obeying $\lambda(1/2)=0$ and $\lambda(1)=1$. The simulation values, plotted alongside $a(p)$ in Figure~\ref{fig:a_lambda}, respect the bound (\ref{eq:rec}) for every $p$: $\lambda(p) \leq a(p)$. The gap $a(p)-\lambda(p) \geq 0$ measures the LIS contribution from non-record fluctuations of the centered walk $W_{k}$ in (\ref{eq:drift}) between consecutive records; we examine this gap in Section~\ref{sec:records:excess}. The gap is non-monotone in $p$, vanishing at both endpoints $p=1/2$ and $p \to 1$ and reaching a maximum of $\simeq 0.1$ near $p \simeq 0.6$.

\subsection{Closed-form record rate $\lambda(p)$}
\label{sec:linear:fwk}

For random walks with finite-variance continuous symmetric jump distributions and constant positive drift, the record-density coefficient $\lambda(p)$ is known explicitly~\cite{gms,wbk,msw}. In the classification of~\cite{msw}, the biased Gaussian random walk lies in Regime~IV, where the mean record number grows linearly, $\langle R_{n}\rangle \sim a_{2}(\mu_{p})\mku n$, with prefactor given by \cite[Eq.~(126)]{msw}
\begin{equation}
\label{eq:lambdagauss}
\begin{split}
\lambda(p) \equiv a_{2}(\mu_{p})
   &= \exp\biggl[-\sum_{n=1}^{\infty}\frac{1}{2n}\operatorname{erfc}\Bigl(\frac{\mu_{p}\sqrt{n}}{\sqrt{2}}\Bigr)\biggr] \\[1ex]
   &= \exp\biggl[-\sum_{n=1}^{\infty} \frac{1}{n}\Phi(-\mu_{p}\sqrt{n})\biggr].
\end{split}
\end{equation}
The same expression admits a renewal-theoretic interpretation. Under positive drift the ascending ladder epochs of the walk form a renewal process with finite mean $\EE(\tau_{+}) = 1/\lambda(p)$, and (\ref{eq:lambdagauss}) is the reciprocal of the Spitzer--Baxter formula \cite{spitzer1956}
\begin{equation}
\EE(\tau_{+}) = \exp\biggl(\sum_{n \geq 1} \frac{1}{n}\PP(X_{n} \leq 0)\biggr)
\end{equation}
for the mean ascending ladder epoch, evaluated for the Gaussian walk with $\PP(X_{n} \leq 0) = \Phi(-\mu_{p}\sqrt{n})$. Two limiting checks are immediate. As $\mu_{p}\to 0^{+}$, $\Phi(-\mu_{p}\sqrt{n}) \to 1/2$ and the sum diverges as $\frac{1}{2}\log{n}$, so $\lambda(p) \to 0$, recovering the Sparre Andersen sublinear regime at the symmetric point. As $\mu_{p}$ grows large, $\Phi(-\mu_{p}\sqrt{n})$ decays super-exponentially in $n$, the sum tends to zero, and $\lambda(p) \to 1$, recovering the near-deterministic limit. The large-$\mu_{p}$ asymptotics of $\lambda(\mu_{p})$ were derived in \cite{wbk},
\begin{equation}
\label{eq:lambdalarge}
\lambda(\mu_{p}) \simeq 1 - \frac{e^{-\mu_{p}^{2}/2}}{\sqrt{2\pi}\mku\mu_{p}} ~\text{ as }~ \mu_{p} \to \infty.
\end{equation}
A Mellin-transform analysis of (\ref{eq:lambdagauss}), given in Appendix~\ref{app:smalldrift}, yields the exact small-drift slope
\begin{equation}
\label{eq:lambdasmall0}
\lambda(\mu_{p}) \simeq \sqrt{2}\mku\mu_{p} ~\text{ as }~ \mu_{p} \downarrow 0,
\end{equation}
consistent with the numerical estimate $\lambda \approx 1.39\mku\mu_{p}$ reported in \cite{wbk}. As a further check on (\ref{eq:lambdagauss}), at $\mu_{p}=1$ we recover the value $\lambda = 0.800543\ldots$ quoted in \cite{msw}.

Equation (\ref{eq:lambdagauss}) is the established Regime IV result of \cite{msw}, not a prediction of this work. We use it here as a theoretical anchor for the record process, against which the new LIS-side observables of Section~\ref{sec:records} can be interpreted. No analogous closed-form prediction for the LIS coefficient $a(p)$ of Section~\ref{sec:linear:ap} is currently available.

The numerical evaluation of (\ref{eq:lambdagauss}) over our values of $\mu_{p}$ is overlaid on the simulation data in Figure~\ref{fig:a_lambda}. The prediction agrees with the sample averages to within $1$--$3$ standard errors of the mean at every point. The residuals $|\widehat{\lambda}(p)-\lambda(p)|$ are at the $10^{-5}$ level for $p \geq 0.52$ and at $10^{-4}$ for $p = 0.51$ and $0.53$. Sample standard errors are $\mathrm{SE}(\widehat{\lambda}) \sim 10^{-5}$ at our $10^{4}$ samples; no systematic deviation is resolved. The record process of the biased Gaussian random walk is thus quantitatively accounted for by (\ref{eq:lambdagauss}).

\subsection{Distribution of $L_{n}$}
\label{sec:linear:dist}

The linear regime also has a distributional signature. For $p>1/2$, the LIS length $L_{n}(p)$ is a sum of order $n$ approximately independent contributions (modulo the global monotonicity constraint), and a central-limit-type argument suggests that the standardized $L_{n}(p)$ should converge to a Gaussian at fixed $p$ for large $n$. The symmetric case $p=1/2$ falls outside the linear regime and the same argument does not apply; for the sublinear-LIS regime, the bulk of the $L_{n}$ distribution is believed to be well-approximated by a lognormal \cite{weakheavy}.

\begin{figure}[t]
\centering
\includegraphics[viewport= 0 10 480 460, scale=0.40, clip]{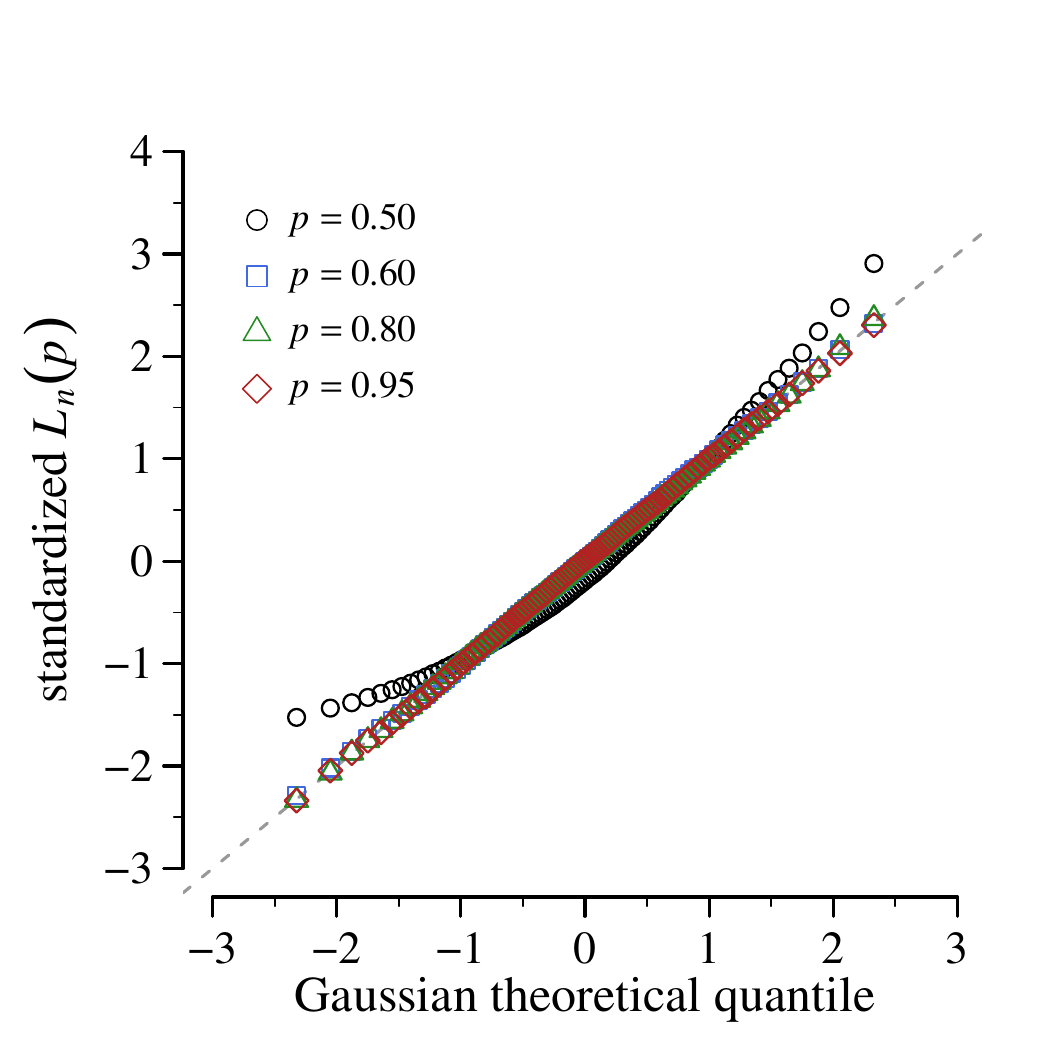}
\caption{Quantile--quantile plot of the standardized LIS length $(L_{n}-\langle L_{n}\rangle)/\operatorname{sd}(L_{n})$ at $n=10^{6}$ against the standard Gaussian quantile. At $p=1/2$, the distribution is right-skewed and fat-tailed, consistent with the lognormal-like bulk reported for symmetric walks. For the sampled $p>1/2$, the points follow the Gaussian reference line within Monte Carlo scatter.}
\label{fig:qq}
\end{figure}

Figure~\ref{fig:qq} shows the quantile-quantile plot of the standardized $L_{n}$ at $n=10^{6}$ against the Gaussian quantile, for four representative values of $p$. At $p=1/2$ the data depart visibly from the $y=x$ reference: the right tail of $L_{n}$ is fatter than Gaussian, with empirical skewness $+0.92$ and excess kurtosis $+0.88$. A Kolmogorov--Smirnov test rejects Gaussianity overwhelmingly ($P\text{-value} \sim 10^{-52}$). The lognormal fit is much closer but also formally rejected at our sample size ($P\text{-value} \sim 10^{-4}$); the bulk-fit interpretation of \cite{weakheavy} should be read in that spirit. The KS $P$-values quoted here are diagnostic: with location and scale estimated from the same sample, the standard KS reference distribution does not strictly apply. The asymmetric cases shown are $p = 0.60$, $0.80$, and $0.95$; the results at $p=0.51$ and at the other sampled values $p>1/2$ are visually indistinguishable from the reference line. The QQ points fall on the diagonal to within Monte Carlo scatter:
\begin{equation}
\begin{split}
\abs{\text{skewness}} &\lesssim 0.02 \quad (\sqrt{6/N} \approx 0.024), \\[1ex]
\abs{\text{excess kurtosis}} &\lesssim 0.07 \quad (\sqrt{24/N} \approx 0.049),
\end{split}
\end{equation}
against the Gaussian sampling standard deviations indicated in parentheses at $N=10^{4}$, and the Kolmogorov--Smirnov test cannot reject Gaussianity at any conventional significance level. The numerical evidence therefore separates the two regimes clearly: lognormal-like at $p=1/2$, and consistent with Gaussian fluctuations for every sampled $p>1/2$, with the central-limit picture already in effect at the smallest asymmetry tested.


\section{The record skeleton and the LIS}
\label{sec:records}

\subsection{Diagnostic ratios}
\label{sec:records:ratios}

\begin{figure}[t]
\centering
\includegraphics[viewport= 0 10 480 460, scale=0.40, clip]{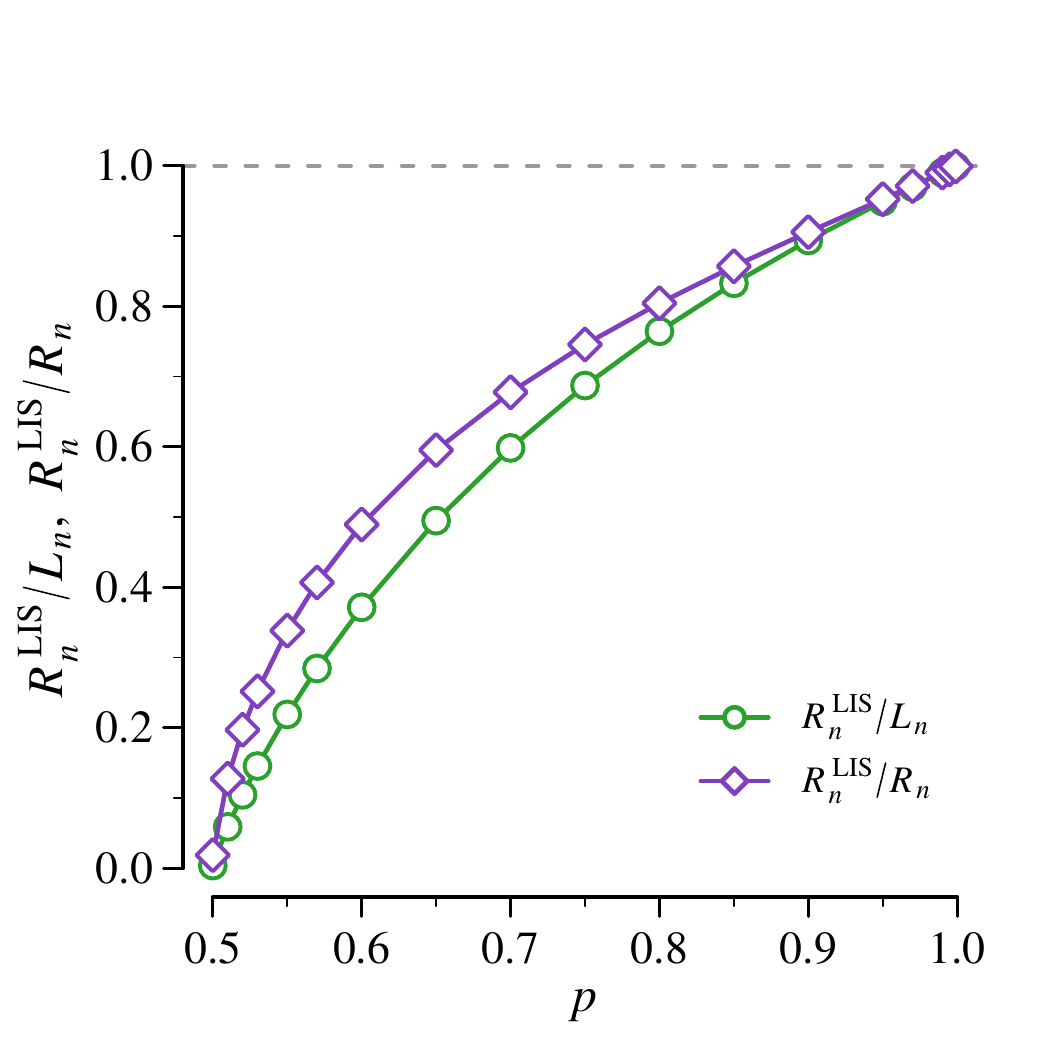}
\caption{Mechanism diagnostic ratios $R_{n}^{\text{LIS}}/L_{n}$ and $R_{n}^{\text{LIS}}/R_{n}$ for the biased Gaussian random walk at $n=10^{7}$. The ratios are small at $p=1/2$ and approach unity as $p\to1$. Their separation in the intermediate range reflects the non-record contribution to the LIS.}
\label{fig:ratios}
\end{figure}

The fraction $R_{n}^{\text{LIS}}/L_{n}$ measures the proportion of LIS elements that are records of the walk and tends to $r(p)/a(p)$ as $n \to \infty$, while the fraction $R_{n}^{\text{LIS}}/R_{n}$ measures the proportion of walk records that participate in a given LIS and tends to $r(p)/\lambda(p)$. At $p=1/2$, the fraction $R_{n}^{\text{LIS}}/L_{n}$ is expected to vanish asymptotically in the symmetric finite-variance regime, since $R_{n}$ grows as $O(\sqrt{n})$ whereas $L_{n}$ has an additional logarithmic factor (\ref{eq:short}). The behavior of $R_{n}^{\text{LIS}}/R_{n}$ at $p=1/2$ depends on which LIS is recovered, since different maximal-length LISes can use different subsets of the $O(\sqrt{n})$ records. In our simulations, the reconstructed LIS uses a small fraction of the records at $p=1/2$, and this fraction rises rapidly once a positive drift is introduced. Both ratios approach unity as $p \to 1$, where the LIS coincides asymptotically with the record skeleton. The two ratios are plotted as functions of $p$ at $n=10^{7}$ in Figure~\ref{fig:ratios}.

By construction, $R_{n}^{\text{LIS}}/R_{n} \geq R_{n}^{\text{LIS}}/L_{n}$ pointwise, since dividing the same intersection by the larger denominator $L_{n} \geq R_{n}$ yields a smaller number, and equivalently $r(p)/\lambda(p) \geq r(p)/a(p)$ in the limit. Between the two endpoints the curves exhibit a visible gap, the same mechanism as the difference $a(p)-\lambda(p)$ of Figure~\ref{fig:a_lambda}: the fraction of records not used by the LIS versus the fraction of LIS elements that are non-records.

\subsection{The fluctuation excess}
\label{sec:records:excess}

The difference $a(p)-\lambda(p)$, non-negative by (\ref{eq:rec}), quantifies the fraction of the walk that participates in the LIS but is not itself a record, that is, the contribution of local fluctuations of the centered walk $W_{k}$ in (\ref{eq:drift}) between consecutive records. The fluctuation excess vanishes at both $p=1/2$, since both $a$ and $\lambda$ vanish there (with a small finite-$n$ residue of order $\sqrt{n}/n$ visible in our $n=10^{7}$ data, see Section~\ref{sec:symlimit}), and $p \to 1$, because both saturate at $1$ together. In between, $a(p)-\lambda(p)$ is a bell-shaped function of $p$, visible in Figure~\ref{fig:a_lambda} as the vertical gap between the two curves; the peak is $\simeq 0.098$, attained at $p=0.60$ ($\mu_{p} \simeq 0.25$), with our resolution placing the true maximum somewhere in $0.57 \leq p \leq 0.65$. At this peak the LIS contains about $10\%$ more of the walk than the record skeleton alone would furnish, the extra elements being non-record steps incorporated between consecutive records. The excess falls to half its peak at $p \simeq 0.53$ on the symmetric side and at $p \simeq 0.75$ on the asymmetric side.

This asymmetry of the bell curve in $p$ is largely intrinsic to $(a-\lambda)$ viewed as a function of the drift $\mu_{p}$: a singular, slower-than-linear rise on the small-drift side (Section~\ref{sec:symlimit:approach}) gives way to a much faster Gaussian-tailed decay on the large-drift side, where $\lambda(\mu_{p}) \to 1$ exponentially fast \cite{wbk}. The bias-to-drift map only mildly modifies this asymmetry in $p$-coordinates,
\begin{equation}
\begin{split}
\mu_{p} &\sim \sqrt{2\pi}\mku(p-1/2) ~\text{ as }~ p \to 1/2, \\[1ex]
\mu_{p} &\sim \sqrt{-2\log(1-p)} ~\text{ as }~ p \to 1.
\end{split}
\end{equation}
Figure~\ref{fig:ratios} gives the corresponding ratio-based view. The vertical gap between the two diagnostic ratios is widest in the same region $0.55 \leq p \leq 0.70$, where the LIS contains its largest mixture of record and non-record elements. Outside this region the two ratios approach each other: near the symmetric point the LIS is fluctuation-driven, while near the deterministic limit it is record-driven. The fluctuation excess is small even at its peak but persists throughout the interior of the $p$ range.


\section{The symmetric singular limit}
\label{sec:symlimit}

The symmetric point $p=1/2$ is a singular limit of the family of biased Gaussian walks studied above. At $p=1/2$ the linear coefficients vanish, $a(1/2)=\lambda(1/2)=0$, and the leading asymptotic behavior of $\langle L_{n} \rangle$ collapses to the symmetric $\sqrt{n}\mku\log{n}$ form (\ref{eq:short}) while $\langle R_{n} \rangle$ collapses to the Sparre Andersen universal $\sqrt{4n/\pi}$ \cite{majumdarziff}. The transition between the symmetric scaling and the linear regime of any $p > 1/2$ is therefore controlled by a drift crossover scale $\mu_{c}(n)$.

\subsection{The critical drift scale}
\label{sec:symlimit:crit}

A simple scaling argument identifies the relevant drift threshold at walk length $n$. The deterministic excursion of the drift term in (\ref{eq:drift}) over $n$ steps is $\mu_{p}\mku n$, while the fluctuations of the centered Gaussian random walk $W_{k}$ over the same range are of order $\sqrt{n}$. The drift dominates the fluctuations when $\mu_{p} \gtrsim n^{-1/2}$, \ie, at the critical drift
\begin{equation}
\label{eq:muc}
\mu_{c}(n) \sim n^{-1/2},
\end{equation}
which, for $n$ in the range studied here, corresponds to $\mu_{c} = 10^{-2}$ and $10^{-3}$ or, equivalently, $p_{c}(n)-1/2 = 4\times 10^{-3}$ and $4\times 10^{-4}$. Our smallest sampled drift $\mu_{0.51} \approx 0.025$ sits well above $\mu_{c}$ at every $n$ we simulate, which is why we observe $\theta_{\text{eff}} \simeq 1$ throughout.

\subsection{Approach to the symmetric point}
\label{sec:symlimit:approach}

For small drift, the record-density coefficient has the closed-form limit
\begin{equation}
\lambda(\mu_{p}) \simeq \sqrt{2}\,\mu_{p} ~\text{ as }~ \mu_{p} \downarrow 0,
\end{equation}
derived from (\ref{eq:lambdagauss}) in Appendix~\ref{app:smalldrift}. No analogous expression is available for the LIS coefficient $a(\mu_{p})$. The relevant small-drift question is therefore the behavior of the fluctuation excess
\begin{equation}
a(\mu_{p})-\lambda(\mu_{p}),
\end{equation}
\ie, whether $a(\mu_{p})$ approaches zero linearly with $\mu_{p}$ or more slowly.

A simple inter-record heuristic suggests a singular lift-off. Since the typical separation between consecutive records is of order $1/\lambda(\mu_{p})\sim 1/\mu_{p}$, one may regard the centered part of the walk between records as contributing fluctuation segments of this typical length. If the symmetric finite-variance estimate $L_{m} \sim \sqrt{m}\mku\log m$ were applied independently to each such segment, then each inter-record interval would contribute of order
\begin{equation}
\mu_{p}^{-1/2}\log\Big(\frac{1}{\mu_{p}}\Big)
\end{equation}
non-record LIS elements. Summing over $R_{n} \sim \mu_{p}\mku n$ intervals would give the naive estimate
\begin{equation}
a(\mu_{p})-\lambda(\mu_{p}) \sim \sqrt{\mu_{p}}\,\log\Big(\frac{1}{\mu_{p}}\Big) ~\text{ as }~ \mu_{p} \downarrow 0.
\end{equation}
This argument over-counts because LIS pieces from distinct inter-record intervals cannot be concatenated freely. The global monotonicity constraint forces each piece to start above the last selected element of the preceding interval, possibly introducing a $\mu_{p}$-dependent suppression.

\begin{figure}[t]
\centering
\includegraphics[viewport= 10 10 520 400, scale=0.475, clip]{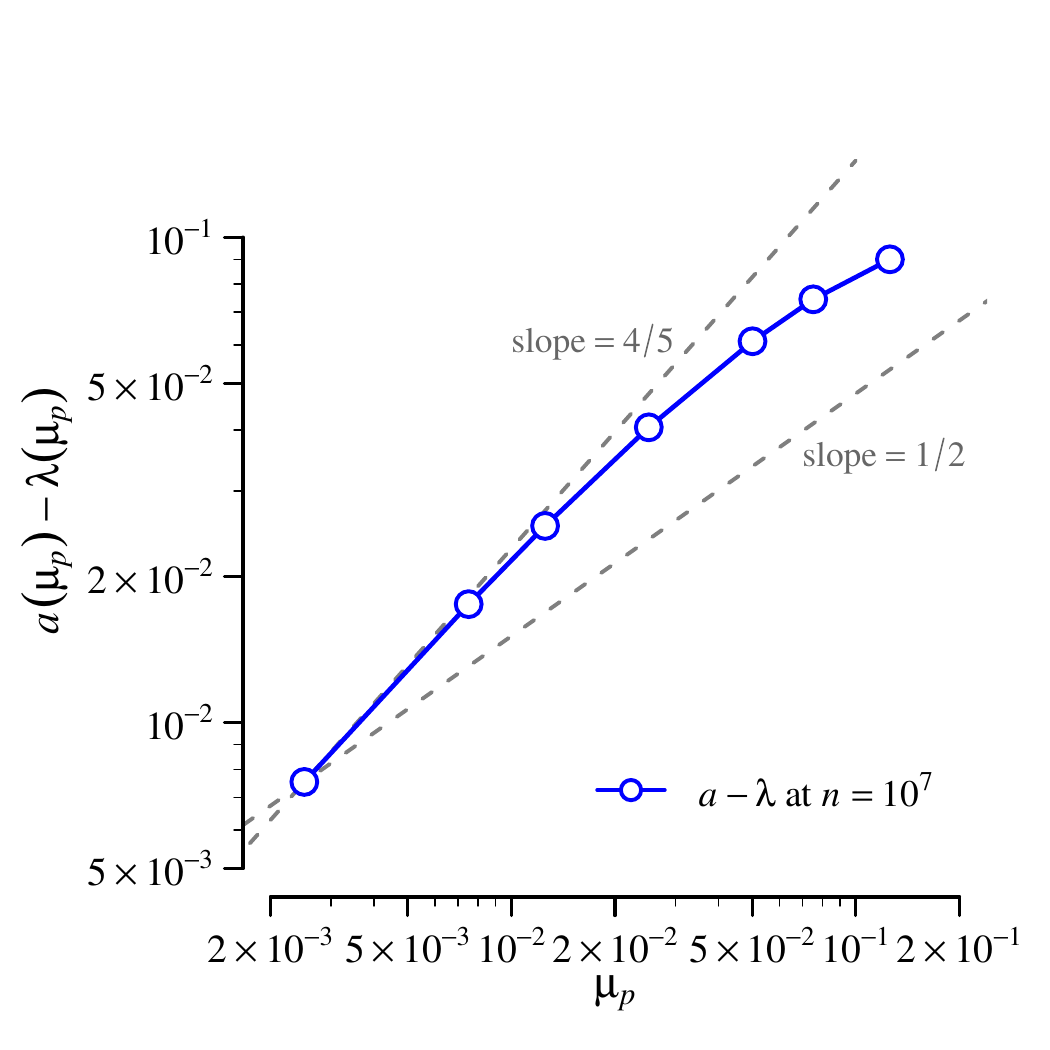}
\caption{Fluctuation excess $a(\mu_{p})-\lambda(\mu_{p})$ versus drift $\mu_{p}$ at $n=10^{7}$ for $p = 0.501, 0.503, 0.505, 0.51, 0.52, 0.53, 0.55$. The dashed guide lines, anchored at the smallest drift, have slopes $1/2$ and $4/5$. The local slope drifts across the sampled range, indicating that no single power law is resolved.}
\label{fig:amla_loglog}
\end{figure}

To test this boundary layer numerically, we extended the simulations to the small-asymmetry points $p = 0.501, 0.503,$ and $0.505$ and to walk lengths up to $n=10^{7}$. Figure~\ref{fig:amla_loglog} shows the resulting fluctuation excess $a(\mu_{p})-\lambda(\mu_{p})$ on a log-log scale. The data do not follow a single power law over the sampled range. A regression over the seven points $p = 0.501, 0.503, 0.505, 0.51, 0.52, 0.53$, and $0.55$ gives an effective slope $0.640 \pm 0.030$, but the local slope drifts from about $0.77$ at the smallest sampled drifts to about $0.37$ near the upper end of the fitting window. Thus the earlier apparent $\sqrt{\mu_{p}}$ behavior was a local effective slope, not an asymptotic law.

Over the sampled range, the data are incompatible with both the bare $\sqrt{\mu_{p}}$ form and the naive $\sqrt{\mu_{p}}\log(1/\mu_{p})$ prediction. They do, however, support the weaker singular conclusion
\begin{equation}
\frac{a(\mu_{p})}{\mu_{p}} \to \infty ~\text{ as }~ \mu_{p} \downarrow 0,
\end{equation}
\ie, that the LIS coefficient vanishes more slowly than linearly in the drift. Quantitatively, $a/\mu_{p}$ grows from approximately $3.01$ at $p=0.51$ to $4.42$ at $p=0.501$, while $a/\lambda$ grows from approximately $2.16$ to $3.12$ over the same range. The precise asymptotic form remains open.

\section{Summary and conclusions}
\label{sec:summary}

We have studied the longest increasing subsequence of the biased Gaussian random walk, with drift parametrized by $p = \PP(\xi>0) \geq 1/2$ or, equivalently, $\mu_{p} = \Phi^{-1}(p)$. For every fixed $p>1/2$, the mean LIS length, the mean record count, and the mean LIS-record overlap grow linearly with the walk length,
\begin{equation}
\begin{split}
\langle L_{n}(p)\rangle &\sim a(p)n, \\[1ex]
\langle R_{n}(p)\rangle &\sim \lambda(p)n, \\[1ex]
\langle R_{n}^{\text{LIS}}(p)\rangle &\sim r(p)n.
\end{split}
\end{equation}
Thus the exponent question familiar from the symmetric and heavy-tailed settings ceases to be informative here; any positive drift gives $\theta(p)=1$, and the nontrivial object is the coefficient $a(p)$. The record coefficient $\lambda(p)$ is quantitatively accounted for by the known record-rate formula~\cite{msw}, while no analogous closed-form expression is presently available for $a(p)$.

The linear regime has a simple mechanism interpretation. The positive drift creates an extensive record skeleton, and the LIS increasingly aligns with this skeleton as $p\to1$. The difference $a(p)-\lambda(p)$ measures the non-record contribution to the LIS: it vanishes at both endpoints, peaks near $p \simeq 0.60$ ($\mu_{p} \simeq 0.25$), and reaches about $10\%$ at its maximum. Near the symmetric point, the limit is singular. The record density satisfies $\lambda(\mu_{p}) \simeq \sqrt{2}\mu_{p}$, whereas the data indicate that $a(\mu_{p})$ vanishes more slowly than linearly: $a(\mu_{p})/\mu_{p}$ increases as $\mu_{p} \downarrow 0$. The local log-log slope of $a(\mu_{p})-\lambda(\mu_{p})$ drifts across the sampled range, so no single power law is resolved; in particular, the naive inter-record excursion estimate does not capture the observed small-drift behavior. The finite-size crossover between the symmetric and linear regimes is controlled by the drift scale $\mu_c(n) \sim n^{-1/2}$.

The distribution of $L_{n}(p)$ also changes across the singular point. At $p=1/2$, the standardized LIS length remains right-skewed and is consistent with the lognormal-like bulk observed in previous work on symmetric walks. For every sampled $p>1/2$, the standardized fluctuations are instead consistent with a Gaussian, in agreement with a central-limit picture for an extensive observable in the linear regime. These observations suggest that $L_{n}/n$ is self-averaging for fixed positive drift.

The main open problem is to derive the LIS coefficient $a(p)$. Any candidate theory must reproduce the lower bound $a(p) \geq \lambda(p)$, the high-drift limit $a(p) \to p$, and the singular small-drift behavior reported here. Such a derivation would have to combine the renewal structure of the record skeleton with the global monotonicity constraint that governs the insertion of non-record LIS elements between records, which appears to be a genuinely hard problem. An infinite-variance counterpart worth exploring is the biased Cauchy random walk, which lies in a different record-statistical regime, where the record exponent itself depends continuously on the drift.


\begin{acknowledgments}
The author thanks FAPESP (Brazil) for partial financial support through research grant no.~2020/04475-7.
\end{acknowledgments}


\appendix

\section{Small-drift limit of the record rate}
\label{app:smalldrift}

The exact Regime~IV prefactor (\ref{eq:lambdagauss}) was given in \cite{msw}. Its small-drift behavior is governed by a result of Chang and Peres \cite{changperes} on the first ascending ladder height of the Gaussian walk. We recover the slope below by a Mellin transform argument in the record-rate variable and use it to replace the numerical estimate $\lambda \approx 1.39\,\mu_{p}$, given in \cite{wbk}, by its exact value. 

With $\Phi(-x) = \tfrac{1}{2}\operatorname{erfc}(x/\sqrt{2})$, write $\lambda(\mu_{p}) = e^{-g(\mu_{p})}$, where
\begin{equation}
\label{eq:gdef}
g(\mu_{p}) = \sum_{n=1}^{\infty}\frac{1}{n}\,\Phi(-\mu_{p}\sqrt{n})
           = \frac{1}{2}\sum_{n=1}^{\infty}\frac{1}{n}\operatorname{erfc}\big(\mu_{p}\sqrt{n/2}\big).
\end{equation}
Using
\begin{equation}
\int_{0}^{\infty} t^{s-1} \operatorname{erfc}(t)\mku dt = \frac{1}{s\sqrt{\pi}}\,\Gamma\Bigl(\frac{s+1}{2}\Bigr)
\end{equation}
for $\operatorname{Re}s>0$, term-by-term integration of (\ref{eq:gdef}) gives
\begin{equation}
\label{eq:Gs}
G(s) = \int_{0}^{\infty}\mu^{s-1}g(\mu)\mku d\mu
     = \frac{2^{s/2-1}}{s\sqrt{\pi}}\,\Gamma\Bigl(\frac{s+1}{2}\Bigr)\zeta\!\Bigl(1+\frac{s}{2}\Bigr),
\end{equation}
analytic for $\operatorname{Re}s>0$, with $\zeta$ the Riemann zeta function. Inverting
\begin{equation}
\label{eq:invmellin}
g(\mu) = \frac{1}{2\pi i}\int_{c-i\infty}^{c+i\infty} G(s)\,\mu^{-s}\mku ds
\end{equation}
along any vertical line $\operatorname{Re}s = c$ in the fundamental strip $c>0$, the small-$\mu$ expansion follows by displacing the contour to the left across the poles of (\ref{eq:Gs}). The rightmost is a double pole at $s=0$, where the factor $1/s$ and the simple pole of $\zeta(1+s/2)$ coincide. With $\zeta(1+s/2) = 2/s + \gamma + O(s)$ and $\Gamma((s+1)/2) = \sqrt{\pi}\,\big[1 + \tfrac{s}{2}\mku\psi(1/2) + O(s^{2})\big]$, where $\psi$ is the digamma function and $\psi(1/2) = -\gamma - 2\log\mku2$, the Euler constant $\gamma$ cancels and
\begin{equation}
G(s) = \frac{1}{s^{2}} - \frac{\log\mku2}{2s} + O(1).
\end{equation}
The residue of $G(s)\mku\mu^{-s}$ at this double pole supplies the leading terms $\log(1/\mu_{p}) - \tfrac{1}{2}\log\mku2$, and the next pole, at $s=-1$, contributes the first correction, with residue $-\zeta(1/2)\mku\mu_{p}/\sqrt{2\pi}$. The further poles are simple and confined to the odd negative integers $s=-3,-5,\dots$, inherited from $\Gamma((s+1)/2)$; at the even integers $G(s)$ is regular. So $g(\mu_{p})$ carries only odd powers of $\mu_{p}$, and the first correction is followed by $O(\mu_{p}^{3})$,
\begin{equation}
\label{eq:gasym}
g(\mu_{p}) = \log\Big(\frac{1}{\mu_{p}}\Big) - \frac{1}{2}\log\mku2
   - \frac{\zeta(1/2)}{\sqrt{2\pi}}\,\mu_{p} + O(\mu_{p}^{3}),
\end{equation}
so that
\begin{equation}
\label{eq:lambdacorr}
\lambda(\mu_{p}) = \sqrt{2}\,\mu_{p}\exp\!\Big[\frac{\zeta(1/2)}{\sqrt{2\pi}}\,\mu_{p} + O(\mu_{p}^{3})\Big]
\end{equation}
as $\mu_{p}\downarrow 0$. The leading coefficient is exactly $\sqrt{2}$. Since $\zeta(1/2)\simeq -1.4604 < 0$, the ratio $\lambda(\mu_{p})/\mu_{p}$ approaches $\sqrt{2}$ from below, linearly in the drift. At the smallest asymmetry we simulate, $\mu_{0.51} \simeq 0.025$, Eq.~(\ref{eq:lambdacorr}) gives $\lambda/\mu_{p} \simeq 1.394$, which accounts for the numerical estimate $\lambda \approx 1.39\,\mu_{p}$ in \cite{wbk}. The same Mellin pair (\ref{eq:Gs})--(\ref{eq:invmellin}), continued across the remaining poles, could be used to reproduce the full series of \cite{changperes}. \\[-2ex]

This expansion in the record-rate is a classical result. Writing $\lambda(\mu_{p}) = 1/\EE(\tau_{+})$ for the reciprocal of the mean ascending ladder epoch, Wald's equation \cite{wald1944} gives $\EE_{\mu_{p}}(X_{\tau_{+}}) = \mu_{p}\mku\EE(\tau_{+}) = \mu_{p}/\lambda(\mu_{p})$ for the mean first ascending ladder height $X_{\tau_{+}}$. Equation (\ref{eq:lambdacorr}) is therefore the reciprocal of the Taylor expansion of $\EE_{\mu_{p}}(X_{\tau_{+}})$ about $\mu_{p}=0$ derived in \cite{changperes}, which gives the complete series and shows that it converges in the disk $|\mu_{p}|<2\sqrt{\pi}$. The leading slope $\sqrt{2}$ is the reciprocal of their driftless value $\EE_{0}(X_{\tau_{+}})=1/\sqrt{2}$ (which equals $\sigma/\sqrt{2}$ for step variance $\sigma^{2}$). On the record-statistics side this ladder-height connection has been made only recently and only at zero drift in \cite{godrecheluck}, which studies the first positive position of a symmetric continuous walk (its first ascending ladder height) and recovers the Gaussian series of \cite{changperes} through the Pollaczek--Spitzer formula. For the biased walk, Eq.~(\ref{eq:lambdacorr}) identifies the exact value $\sqrt{2}$ behind the numerical estimate of \cite{wbk}.


\end{document}